**Title:** Use of the Deep Learning Approach to Measure Alveolar Bone Level
**Running title:** Deep Learning Aided Bone Measurement

Chun-Teh Lee[1], DDS, MS, DMSc, Tanjida Kabir[2], MS, Jiman Nelson[1], DMD, MS, Sally Sheng[1], DMD, MS, Hsiu-Wan Meng[1], DDS, MS, Thomas E. Van Dyke[3,4], DMD, PhD, Muhammad F. Walji[5], MS, PhD, Xiaoqian Jiang[2], PhD, Shayan Shams[2,6], PhD

[1] Department of Periodontics and Dental Hygiene, The University of Texas Health Science Center at Houston School of Dentistry, Houston, Texas, USA
[2] The University of Texas Health Science Center at Houston, School of Biomedical Informatics, Houston, Texas, USA
[3] Center for Clinical and Translational Research, The Forsyth Institute, Cambridge, MA, USA
[4] Department of Oral Medicine, Infection, and Immunity, Faculty of Medicine, Harvard University, Boston, MA, USA
[5] Department of Diagnostic and Biomedical Sciences, The University of Texas Health Science Center at Houston School of Dentistry, Houston, Texas, USA
[6] Department of Applied Data Science, San Jose State University, San Jose, California, USA

*The first two authors contributed equally

**Corresponding author:** Shayan Shams
**Address:** One Washington Sq, San Jose, CA 95192
**Tel:** 408-924-2639
**E-mail:** shayan.shams@sjsu.edu

**Word count:** 3485
**Number of figures:** 4**; tables:** 2**; references:** 34

**Acknowledgements:** We would like to thank Mr. Luyao Chen, UTHealth School of Biomedical Informatics, and Mr. Krishna Kumar Kookal, UTHealth School of Dentistry, for retrieving and organizing clinical data and radiographic images.

**Conflict of Interest and Source of Funding:** CL is partially supported by the American Academy of Periodontology Sunstar Innovation Grant. XJ is supported in part by Christopher Sarofim Family Professorship, UT Stars award, UTHealth startup, and the National Institute of Health (NIH) under award number U01 TR002062. SS is supported in part by CPRIT (RR180012, RP200526) NIH (3R41HG010978) grants.

**Abstract**

**Aim**

The goal was to use a Deep Convolutional Neural Network to measure the radiographic alveolar bone level to aid periodontal diagnosis.

**Material and methods**

A Deep Learning (DL) model was developed by integrating three segmentation networks (bone area, tooth, cementoenamel junction) and image analysis to measure the radiographic bone level and assign radiographic bone loss (RBL) stages. The percentage of RBL was calculated to determine the stage of RBL for each tooth. A provisional periodontal diagnosis was assigned using the 2018 periodontitis classification. RBL percentage, staging, and presumptive diagnosis were compared to the measurements and diagnoses made by the independent examiners.

**Results**

The average Dice Similarity Coefficient (DSC) for segmentation was over 0.91. There was no significant difference in RBL percentage measurements determined by DL and examiners ($p = 0.65$). The Area Under the Receiver Operating Characteristics Curve of RBL stage assignment for stage I, II and III was 0.89, 0.90 and 0.90, respectively. The accuracy of the case diagnosis was 0.85.

**Conclusion**

The proposed DL model provides reliable RBL measurements and image-based periodontal diagnosis using periapical radiographic images. However, this model has to be further optimized and validated by a larger number of images to facilitate its application.

**Clinical Relevance**

**Scientific rationale:** Assessing radiographic bone level is important for periodontal diagnosis. The interpretation of radiographic images is subjective, and accuracy depends on a clinician's experience and knowledge. Artificial intelligence and image analysis can improve reliability.

**Principal findings:** The proposed Deep Learning based aided diagnosis can reliably assess the radiographic bone level and assign bone loss stage.

**Practical implications:** The proposed model can assist the clinician in making an accurate periodontal diagnosis. Furthermore, it is useful for the review of a large number of intraoral radiographic images for quality control of clinical diagnosis and research purposes related to periodontal diseases.

**Introduction**

Periodontitis is a biofilm induced chronic inflammatory disease that is characterized by gingival inflammation and alveolar bone loss around teeth. According to the 2009-2014 National Health and Nutrition Examination Survey (NHANES), approximately 61 million adults over 30 years-old have periodontitis (42.2%) with 7.8% having severe periodontitis (Eke et al., 2018). The latest 2018 periodontitis staging, and grading classification was designed to allow clinicians to assess periodontitis severity, complexity, extent, as well as progression rate, and determine the patient's potential response to treatment (Papapanou et al., 2018). Disease severity can be quantified as clinical attachment loss, alveolar bone loss, or the number of teeth lost. The primary criteria for periodontitis grading are direct or indirect evidence of disease progression. Direct evidence consists of longitudinal documentation of progressive attachment loss and/or radiographic bone loss (RBL). When direct evidence is unavailable, the estimated rate of progression, as indirect evidence, can be quantified as RBL at the most severely affected site in relation to the patient's age (Kornman & Papapanou, 2020; Tonetti, Greenwell, & Kornman, 2018). Measuring RBL is important for making a proper periodontal diagnosis, especially when comprehensive and longitudinal periodontal charting is unavailable. Accurate interpretation of radiographs is important, but clinicians may have different interpretations depending on individual experience and knowledge. Developing a tool to assist clinicians in interpreting and measuring alveolar bone will aid an accurate and reliable periodontal diagnosis.

Deep Learning (DL) models have been extensively utilized in different medical domains, including identifying anatomic structures and detecting pathologic findings on radiographic images (Esteva et al., 2019; Giger, 2018; Shams et al., 2018). Recently, DL-based Computer-Aided Diagnosis (CAD) in oral imaging was developed, but adoption has been limited (Schwendicke, Golla, Dreher, & Krois, 2019). There are few studies measuring alveolar bone level on panoramic radiographs using DL models (Chang et al., 2020; Kim, Lee, Song, & Jung, 2019; Krois et al., 2019; Lee, Kim, Jeong, & Choi, 2018). Panoramic images provide a quick overview of the dentition, but the unignorable distortion and a lack of detail prevent accurate and precise diagnosis of periodontitis and other oral diseases (Akesson, Hakansson, & Rohlin, 1992; Hellen-Halme, Lith, & Shi, 2020; Pepelassi & Diamanti-Kipioti, 1997). The current standard is a visual assessment of intraoral radiographs, which is subject to error. State-of-the-art DL models can offer an objective method for a reliable periodontal diagnosis. The goal of this study was to introduce a novel DL-based CAD model and compare it to clinicians' assessment based on periapical radiographs.

**Materials and Methods**

*Model development*

The study was conducted in accordance with the guidelines of the World Medical Association's Declaration of Helsinki, a study checklist for artificial intelligence in dental research (Schwendicke et al., 2021), and approved by the University of Texas Health Science Center at Houston (UTHealth) Committee for the Protection of Human Subjects (HSC-DB-20-1340). A DL-based CAD model was developed for alveolar bone level assessment and periodontal diagnosis based on intraoral radiographs. All digital intraoral images were taken with a standard film positioning holder and approved by the radiology technicians or radiologists. Convolutional neural networks (CNNs), a neural network-based algorithm designed to process data that exhibits natural spatial invariance, were utilized (LeCun, Bengio, & Hinton, 2015). The proposed model was designed specifically to provide high classification accuracy while maintaining interpretability. It integrated three segmentation networks (bone area, tooth and cementoenamel junction (CEJ)) and image analysis (Figure 1).

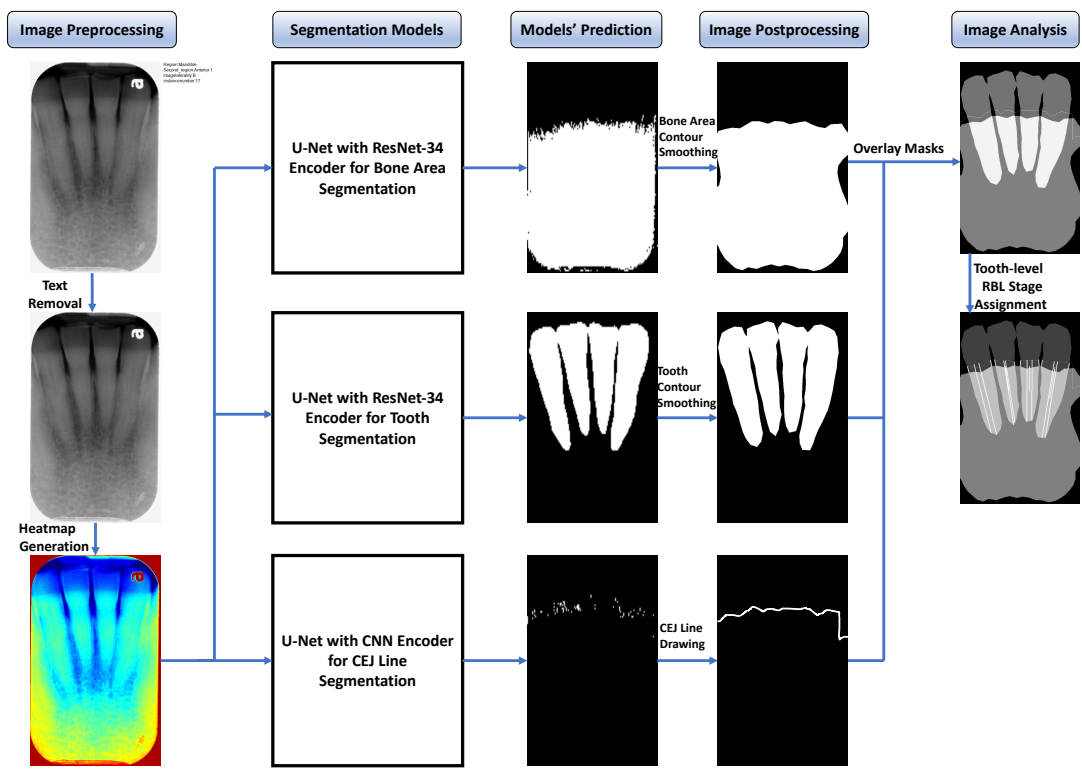

**Figure 1. Flow diagram of the proposed Computer-Aided Diagnosis (CAD) model.** Segmentation models predicted the bone area, teeth and cementoenamel junction (CEJ) masks. Masks are processed to remove noises then overlaid to extract bone area, teeth and CEJ line for radiographic bone loss (RBL) measurement and stage assignment for each tooth.

*Annotation*

An open-source tool, Computer Vision Annotation Tool (CVAT) was used to annotate Region of Interests (ROIs) from each intraoral radiograph. The patients' images were selected from the electronic health record (EHR) system by the periodontal diagnosis history, and the images were extracted from the image database. The extracted Digital Imaging and Communications in Medicine (DICOM) files were converted to Portable Network Graphics (PNG) format with additional metadata such as region and image laterality. Each full mouth series (FMS) radiograph was uploaded to the CVAT under a unique ID for annotation. Multiple ROIs including bone area, tooth, and others (e.g., restorations, vertical defects) were annotated using polygon and CEJ was annotated using polyline on each image via the coordinates and names. When the CEJ was fully covered by the crown and could not be recognized, the margin of the crown was annotated as CEJ. Each clinical examiner received a unique account. They could log in and perform the annotation without observing other examiners' annotations.

*Segmentation model*

Different variations of U-Net (Ronneberger, Fischer, & Brox, 2015) were trained and evaluated to find the best architecture and hyperparameter settings for each segmentation model. Specifically, U-Net with CNN, ResNet-34, and ResNet-50 encoder (He, Zhang, Ren, & Sun, 2016) were evaluated for the bone area, tooth, and CEJ line segmentations. Additionally, the hyperparameters such as kernel size and the number of CNN layers were adjusted to optimize the performance of the model.

The U-Net architecture consists of three paths; the contraction path is composed of a contraction block (CNN) to shrink the input's width and height and double the number of feature maps; the expansion path consists of CNN and Upsampling layers. After each expansion block, the number of feature maps halves while their width and height double to maintain the symmetry; the cross-connection path appends the feature maps of the expansion block to feature maps of corresponding contraction blocks to ensure that features learned during contraction are used in the reconstruction.

U-Net with ResNet-34 encoder (Supplementary Table S1) outperformed other variations for the bone area and tooth segmentations and U-Net with CNN blocks (Supplementary Table S2) provided the best result for the CEJ line segmentation (Table 1).

**Table 1: Performance evaluation matrices of different models for segmenting bone area, tooth and cementoenamel junction (CEJ) line**

| Task Name | Model Name | Pixel Accuracy | Dice Similarity Coefficient | Jaccard Index |
|---|---|---|---|---|
| Segmenting bone area | U-Net | 0.9570 | 0.9621 | 0.9319 |
| | U-Net with ResNet-34 Encoder | 0.9603 | 0.9635 | 0.9343 |
| | U-Net with ResNet-50 Encoder | 0.9610 | 0.9229 | 0.9333 |
| Segmenting tooth | U-Net | 0.8413 | 0.8994 | 0.8225 |
| | U-Net with ResNet-34 Encoder | 0.8660 | 0.9470 | 0.9143 |
| | U-Net with ResNet-50 Encoder | 0.8898 | 0.9533 | 0.9026 |
| Segmenting CEJ line | U-Net with 3X3 kernel | 0.9456 | 0.4287 | 0.2753 |
| | U-Net with 5X5 kernel | 0.9966 | 0.6719 | 0.5143 |
| | U-Net with 7X7 kernel | 0.9850 | 0.9129 | 0.8776 |

The model was trained for 100 epochs with batch-size of 8 using binary-cross entropy loss and the Adam optimizer. Equation (1) illustrates the formula to calculate binary-cross entropy loss:

$$Loss = -\left(\frac{1}{|B|}\right) \sum_{m=1}^{b} \sum_{i=1}^{n} (y_i^m \log \hat{y}_i^m + (1 - y_i^m) \log (1 - \hat{y}_i^m)) \quad (1)$$

Where $\hat{y}_i^m$ denotes the predicted value for pixel $i$ for sample $m$ and $\hat{y}_i^m$ is the target pixel value in the mask, and $B$ is the batch size. The segmentation models' outputs were integrated and utilized for further image analysis to measure alveolar bone level. The model was developed using TensorFlow version 2.0 (Abadi et al., 2016), trained and evaluated using NVIDIA Tesla V100 GPU. The average inference time for a radiographic image is 53.4ms. The source code and data for the project can be obtained by contacting the corresponding author.

*Alveolar bone level measurement*

A sequence of post-processing steps such as noise removal (Gaussian filtering) and precise contour detection was performed to improve the quality of the predicted mask (Figure 2).

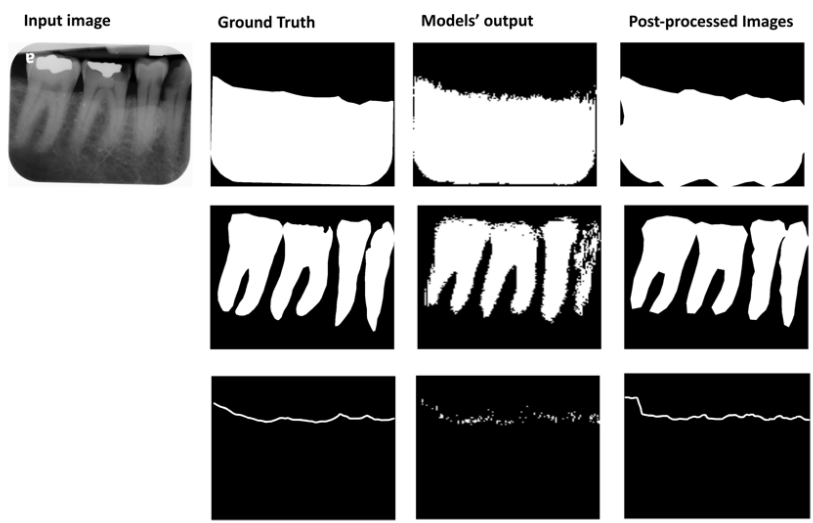

**Figure 2: Input image, ground truth (bone area, tooth and cementoenamel junction (CEJ) line), segmentation models' output, and images after post-processing.**

The parameters in Equation 2, used to calculate RBL percentage, obtained by the following steps:

a) Intersecting points of the bone area and each tooth were obtained by performing logical "AND" operation between the bone area contour and tooth masks (Figure 3A).
b) Intersecting points of the CEJ line and each tooth were obtained by performing logical "AND" operation between the predicted CEJ line and tooth masks (Figure 3B).
c) The minimum points on the left and right sides of each tooth axis were identified to get the apices of each root (Figure 3C).

d) Line 1L and line 1R were drawn from the intersecting points of the tooth and CEJ line (Figure 3D).
e) Line 2L and line 2R were drawn from each root apex to the CEJ line parallel to the tooth axis (Figure 3E).
f) The lines 1L, 2L, 1R, and 2R were transformed to millimeters by multiplying this distance with Imager Pixel Spacing from the DICOM header. The RBL percentage could be obtained (Figure 3F).

$$RBL\% = \max\left(\frac{1L}{2L}, \frac{1R}{2R}\right) \times 100 \qquad (2)$$

Following the 2018 periodontitis classification, stage I: RBL < 15% (in the coronal third of the root), stage II: 15% ≤ RBL ≤ 33% (in the coronal third of the root), stage III: extending to the middle third of root and beyond (RBL > 33%). No bone loss (stage 0) was assigned if the distance between the CEJ and alveolar bone level is less than 1.5 mm (Hausmann, Allen, & Clerehugh, 1991; Xie, 1991) disregarding the RBL percentage. The RBL stage for each tooth was assigned based on the higher RBL stage of the mesial site or the distal site when the two sites have different stages.

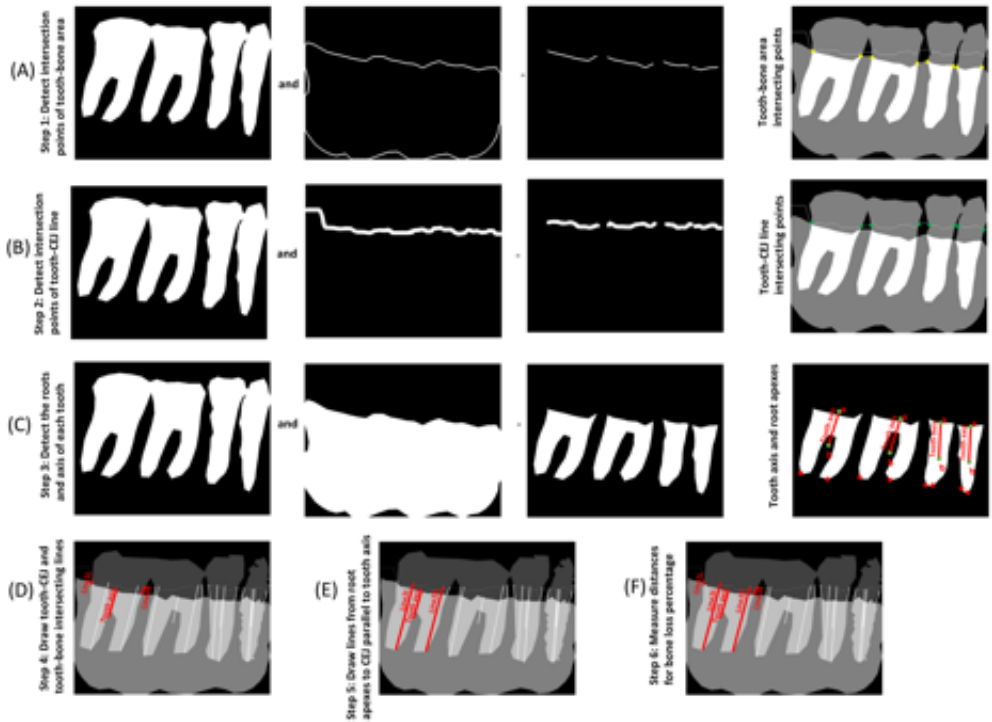

**Figure 3: Image analysis steps for calculating bone loss percentage for each tooth.** Steps to calculate RBL percentage are as follows: (A) identify the intersecting points (yellow dots) between the tooth and bone area; (B) identify the intersecting points (green dots) between the tooth and CEJ line; (C) identify the mesial and distal root axes parallel to the tooth axis to locate the roots' apexes (red dots); (D) calculate the distance between the CEJ and alveolar bone level at both mesial and distal sites (the line connecting the green dots and the yellow dots); (E) calculate the root length by identifying a line from each root apex to the CEJ line parallel to the tooth axis. (F) divide the distance between the CEJ and alveolar bone level by the distance between the CEJ and root apex.

*Model training and validation*

In total, 693 periapical radiographic images from randomly selected 37 periodontitis patients were included in the original dataset. All images were annotated and examined by three independent examiners, including two periodontists and one periodontal resident (S.S, H.M, J.N) familiar with 2018 periodontitis classification. The image dataset was randomly divided to 70%, 10%, and 20% for training, validation, and testing. The model was further evaluated on 644 additional periapical images ("additional dataset") from randomly selected 46 cases to assess the accuracy of RBL percentage measurement, RBL stage assignment, and whole-case periodontal diagnosis. These patients included in the "additional dataset" and the original dataset were completely different to avoid any data snooping bias.

All examiners were calibrated for annotation and RBL percentage measurements using three sets of FMS intraoral radiographs. Before starting the annotations, the Dice Similarity Coefficient (DSC) for annotations between examiners reached at least 0.84, and RBL percentage results between examiners were not significantly different ($p > 0.05$ calculated by Student's t-test). The examiners used the software (MiPACS, Medicor Imaging, Charlotte, NC, USA) to measure RBL percentage, and the time was documented. The segmentation models were trained using the annotated bone area, teeth, and the CEJ line. The stages of RBL were assigned by the examiner before measuring the RBL percentage to avoid potential biases. If there was a conflict among examiners' RBL stage assignment, the stage was decided by the measured RBL percentage to get the ground truth for the final stage assignment.

Forty-six cases' FMS periapical radiographs ("additional dataset") were analyzed by the DL model to assign a periodontal diagnosis for individual cases following the 2018 periodontitis classification. A periodontitis case should have interdental RBL detectable at ≥ 2 non-adjacent teeth. In each case, the RBL percentage or stage of each tooth was assigned based on the highest when a tooth is present in more than one image. The model-assigned periodontal diagnosis (extent, stage, and grade) was based on the proportion of RBL stages, RBL percentages of the teeth and the patient's age. These cases were also diagnosed by the three examiners following the same principle. If there were inconsistent results among the examiners, a diagnosis would be decided by the majority rule.

*Statistical analysis*

The segmentation models were validated by DSC, Jaccard Index (JI), and Pixel Accuracy. DSC (Equation 3) compares the similarity of model's output against reference masks. JI (Equation 4) measures the similarity and diversity of the model's prediction and reference masks. Pixel accuracy (Equation 5) illustrates the percentage of images that are correctly classified by the model with respect to the reference images.

$$DSC = \frac{(2 * the\ Area\ of\ Overlap)}{(Total\ number\ of\ pixels\ in\ both\ images)} \qquad (3)$$

$$Jaccard\ Index = \frac{(Area\ of\ Overlap)}{(Area\ of\ Union)} \qquad (4)$$

$$Pixel\ Accuracy = \frac{(Number\ of\ correctly\ classified\ pixels)}{(Total\ number\ of\ pixels)} \qquad (5)$$

The results of RBL stage assignment by the model were evaluated by the Area Under the Receiving Operating Characteristics Curve (AUROC), sensitivity, specificity, and accuracy. The RBL percentage measurement and required time were compared to examiners' results by Student's t-test. The agreement of RBL stage assignment between the CAD and examiners was calculated by Cohen's Kappa coefficient. The case diagnosis accuracy was presented as the number of cases where the CAD assigned diagnosis is consistent with the examiners divided by the total number of cases. When validating the model assigned RBL stage, an RBL percentage variation (±3 %) was considered in all stage assignments.

Assuming a standard deviation of 0.15, power analysis was conducted for a two-sided t-test with a null hypothesis that the difference between means of the RBL percentages (between different examiners) is no larger than 0.03 at a significance level of 0.05 and a power of 0.8. The analysis showed that at least 394 samples were required.

**Results**

*Segmentation model validation*

For the bone area, tooth shape and CEJ line segmentations, DSC was 0.96, 0.95 and 0.91, JI was 0.93, 0.91 and 0.88, and Pixel Accuracy was 0.96, 0.89, 0.99, respectively.

*Staging validation*

The proposed CAD achieved high AUROC, sensitivity, specificity, and accuracy for RBL stage assignment. While analyzing the "additional dataset", the AUROC values of RBL stage assignment for no bone loss, stage I, stage II, and stage III were 0.98, 0.89, 0.90 and 0.90, respectively. Sensitivity, specificity, and accuracy for different stages are all over 0.8 (Table 2). The RBL stage assignment between the proposed CAD and the ground truth showed higher agreement ($\mathcal{K} = 0.81$) compared to the assignment between individual examiners and the ground truth (examiner 1: $\mathcal{K} = 0.72$; examiner 2: $\mathcal{K} = 0.66$; examiner 3: $\mathcal{K} = 0.52$).

**Table 2. Radiographic bone loss (RBL) stage assignment performance.** AUROC: the Area Under the Receiving Operating Characteristics Curve; sensitivity: true positive/ (true positive + false negative); specificity: true negative/ (true negative + false positive); accuracy: (true positive + true negative)/ (true positive + true negative + false positive + false negative).

| Performance | AUROC | Sensitivity | Specificity | Accuracy |
|---|---|---|---|---|
| Stage 1 RBL | 0.89 | 0.82 | 0.97 | 0.91 |
| Stage 2 RBL | 0.90 | 0.93 | 0.86 | 0.88 |
| Stage 3 RBL | 0.90 | 0.80 | 0.99 | 0.99 |
| No bone loss | 0.98 | 0.96 | 1.00 | 0.99 |

*RBL percentage and level measurements*

There was no significant difference in RBL percentage measurements between the DL model and examiners ($p = 0.65$, Figure 4) in analyzing the "additional dataset". However, the time required to complete RBL percentage measurements and data entry for each FMS radiograph by examiners (mean ± standard deviation: 137 ± 62 minutes) was significantly longer than the time required by the CAD (9 ± 0.51 minutes).

The mean RBL percentages for stage I, II, III measured by the model were 12.11 ± 2.84%, 21.04 ± 2.67 % and 36.12 ± 7.84%, respectively. The mean distances between CEJ and alveolar bone level for sites with RBL stage I, II, III estimated using CAD were 1.73 ± 0.48 mm, 2.93 ± 0.50 mm, and 4.71 ± 1.37 mm, respectively. In the sites assigned with no bone loss, the mean distance was 1.46 ± 0.43 mm.

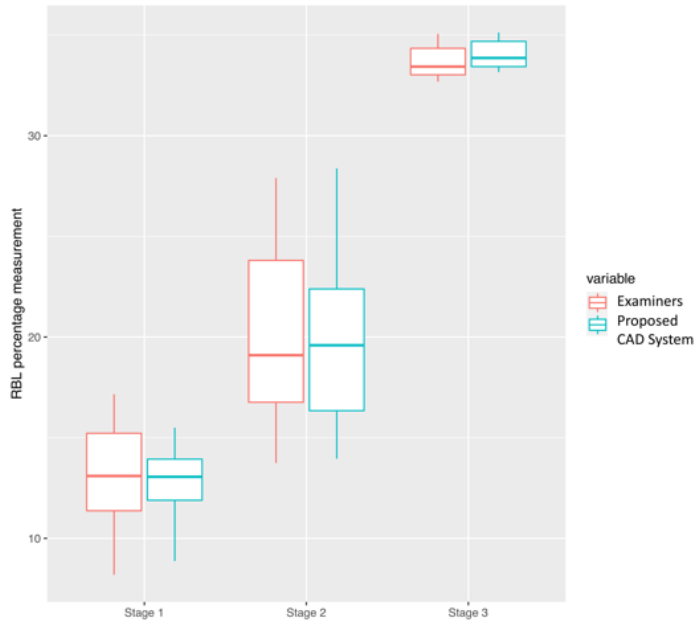

**Figure 4. Radiographic bone loss (RBL) percentage measurement distribution of the proposed Computer-Aided Diagnosis (CAD) model and examiners.** There was no significant difference in RBL percentage measurements between the CAD and examiners (p-value for all cases, stage I cases, stage II cases, and stage III cases= 0.65, 0.32, 0.27, 0.96). The bar inside the box represents the median. The upper end of the box represents the third quartile, and the lower end of the box represents the first quartile. The ends of the whiskers represent maximum and minimum.

*Periodontitis diagnosis assignment*

The 46 cases from the "additional dataset" for diagnosis testing included 16 cases with localized, stage III grade B diagnoses, 10 cases with localized, stage III, grade C diagnoses, 10 cases with generalized, stage III, grade C diagnoses and cases with other diagnoses (Supplementary Table S3). The accuracy of periodontitis case diagnosis was 0.85 in the 46 cases. The accuracy for extent (0.96), stage (0.87) and grade (0.96) were generally high.

**Discussion**

The proposed DL-based CAD model demonstrates high accuracy of alveolar bone level measurements and provisional periodontitis diagnosis based on periapical radiographs. Although, ideally, periodontal diagnosis should be made based on periodontal examinations, radiographic images, and clinical judgement together, the DL model using radiographic findings can provide a rapid and reliable preliminary periodontal diagnosis. Notably, the CAD analysis was considerably faster than clinicians (9 vs. 137 minutes). The proposed DL model is helpful when a comprehensive periodontal examination is missing, and the inexperienced clinician cannot make a proper periodontal diagnosis, or a large number of patients' periodontal diagnosis and radiographic findings have to be re-assessed for diagnosis quality control and research purposes.

Few studies have investigated the utilization of DL in assessing the alveolar bone level and periodontitis staging. Lee et al. developed a DL model to classify periodontally compromised teeth from periapical radiographs (Lee et al., 2018). Kim et al. and Krois et al. utilized DL to detect RBL or calculate RBL percentage from panoramic radiographs (Kim et al., 2019; Krois et al., 2019). However, these models did not assign RBL stage to each tooth or determine the case-level periodontitis stage and grade. Chang et al. developed a hybrid model using image processing and DL to assign periodontitis stage from panoramic radiographs (Chang et al., 2020). However, the AUC for RBL stage assignment and the whole-case diagnosis accuracy were not provided. Comparing the current results to the results in the literature, the segmentation accuracy of the proposed model was comparable with or superior to the other study whose DSC, JI and Pixel Accuracy ranged from 0.83 to 0.93 (Chang et al., 2020). The current model also achieved a comparable or higher accuracy of detecting different bone loss levels as compared to the other studies (Kim et al., 2019; Krois et al., 2019). It should be noted that those published studies analyzed panoramic radiographs. The proposed model is the only model measuring RBL and assigning periodontitis extent, stage, and grade compatible with the 2018 periodontitis classification based on periapical radiographs, the standard radiographic images for periodontal diagnosis (Do, Takei, & Carranza, 2018; American Academy of Periodontology, 2003; Tetradis, Mallya, & Takei, 2018). Although the accuracy and reliability of the other models appear to be good, assessing bone level on panoramic radiographs to make periodontitis diagnosis is generally not recommended due to distorted images, overlapping objects, and low resolution.

The current results showed that clinicians do not assign bone loss stage accurately without manually calculating RBL percentage for each tooth, a very time-consuming process, suggesting a CAD tool designed to assess RBL will be helpful in clinical decision-making. However, RBL percentage can be misleading in some clinical situations, such as teeth with short roots or wide supracrustal tissue attachment. Our DL model also provides information on the distance between CEJ and alveolar bone level, which was not reported in any published DL model, for the clinician's reference. The average CEJ-bone distance for no bone loss sites in this study was similar to the results in the literature (Hausmann et al., 1991; Xie, 1991) and close to the distance for sites with stage I bone loss. These findings support the concept that initial radiographic bone loss should be diagnosed based on multiple factors, including shape of the alveolar bone crest, discontinuity of the crestal lamina dura, bone level relative to other teeth, and longitudinal radiographic bone level change (Zaki, Hoffmann, Hausmann, & Scannapieco, 2015).

The accuracy of the whole dentition periodontal diagnosis suggested by the model was generally high. As compared to extent and grade, the accuracy of stage was lower. In this exercise, stage is decided by the teeth with the more severe RBL than others (Kornman & Papapanou, 2020). The current DL model tends to underestimate RBL at vertical defects, endo-periodontal lesions, and at highly overlapped teeth, where defining bone level is more challenging. This could explain the lower accuracy of stage assignment in the diagnosis.

The clinical attachment level (CAL) change is considered the primary parameter used to define the severity of periodontitis, because the periodontal destruction has to be progressive to a certain degree before it can be visualized in the radiograph (Goodson, Haffajee, & Socransky, 1984). Use of RBL to determine the periodontal diagnosis may result in under-detection of incipient periodontitis and underestimate of disease severity, although CAL, very sensitive to the degree of inflammation present, is moderately to highly associated with radiographic bone level (Farook et al., 2020; Zhang, Rajani, & Wang, 2018). However, CAL is not commonly documented in clinical practice (Patel et al., 2020), since it is less relevant to clinical treatment decisions compared to probing depth and level of bone loss. Additionally, the accuracy of CAL is questioned for inexperienced clinicians (Vandana & Gupta, 2009). Although RBL cannot fully represent periodontal destruction, assessing RBL might be a more practical approach to determine periodontal disease severity than measuring CAL in many clinical settings.

In addition to assisting clinical interpretation of RBL, the model can be useful when a large number of radiographic images have to be reviewed. In a teaching clinic, the imaging diagnosis accuracy and reliability are a primary component of quality assurance. The proposed DL model can efficiently review bone level interpretation results of many students. To study periodontal disease progression and treatment outcomes in a large cohort, retrospective review of periodontal charting is frequently performed. However, inconsistency of periodontal charting between different care providers may affect data reliability (Buduneli, Aksoy, Kose, & Atilla, 2004; Lafzi, Mohammadi, Eskandari, & Pourkhamneh, 2007). Bone level changes in periapical images reviewed by the proposed DL model can be used to validate results of periodontal charting (Machtei, Hausmann, Grossi, Dunford, & Genco, 1997; Zaki et al., 2015).

Although the proposed model demonstrated high accuracy and reliability in measuring RBL and assigning a periodontitis diagnosis, it has some limitations. This model is not able to precisely identify vertical defect depth as well as angulation that can be important for periodontal diagnosis in some cases. Images with poor quality, such as overlapping teeth and distorted tooth length that may mislead the diagnosis, cannot be automatically excluded from the analysis. When multiple teeth are missing, the model is not able to accurately identify tooth number (position). To assess tooth-level or case-level RBL of a large cohort, the current model has to be trained and validated by more images to further optimize its performance and efficiency. Additionally, the model is not designed to replace periodontal charting and other clinical data. An accurate periodontal diagnosis should always be made based on the results of the periodontal charting, radiographic findings, and patient history.

With the limitations of this study, the DL model provides an accurate and reliable alveolar bone level measurement, RBL stage assignment and preliminary periodontal diagnosis based on periapical radiographs. DL can be utilized as a tool to assist clinicians in diagnosing periodontitis in the clinic and further making the proper treatment plan.

# Appendices

**Supplementary Table S1. Convolutional kernel sizes of the U-Net with ResNet-34 encoder.**

| U-Net with ResNet-34 encoder | |
|---|---|
| **Encoder** | **Decoder** |
| **Layer Name, Kernel Size, Number of Kernels** | **Layer Name, Kernel Size, Number of Kernels** |
| Conv E_1, $[3 \times 3, \ 64] \times 1$ <br> Max Pooling, $[3 \times 3, \ 1] \times 1$ <br> Conv E_2, $\begin{bmatrix} 3 \times 3, \ 64 \\ 3 \times 3, \ 64 \end{bmatrix} \times 3$ <br> Conv E_3, $\begin{bmatrix} 3 \times 3, \ 128 \\ 3 \times 3, \ 128 \end{bmatrix} \times 4$ <br> Conv E_4, $\begin{bmatrix} 3 \times 3, \ 256 \\ 3 \times 3, \ 256 \end{bmatrix} \times 6$ <br> Conv E_5, $\begin{bmatrix} 3 \times 3, \ 512 \\ 3 \times 3, \ 512 \end{bmatrix} \times 3$ | Upsampling D_1, $[2 \times 2] \times 1$ <br> Conv D_1, $[3 \times 3, \ 256] \times 2$ <br> Upsampling, D_2, $[2 \times 2] \times 1$ <br> Conv D_2, $[3 \times 3, \ 128] \times 2$ <br> Upsampling, D_3, $[2 \times 2] \times 1$ <br> Conv D_3, $[3 \times 3, \ 64] \times 2$ <br> Upsampling, D_4, $[2 \times 2] \times 1$ <br> Conv D_4, $[3 \times 3, \ 32] \times 2$ <br> Upsampling, D_5, $[2 \times 2] \times 1$ <br> Conv D_5, $[3 \times 3, \ 16] \times 2$ <br> CNN + Sigmoid, $[3 \times 3, 1] \times 1$ |

*Conv = CNN + Batch Normalization + ReLU

**Supplementary Table S2. Convolutional kernel sizes of the U-Net.**

| U-Net | |
|---|---|
| **Encoder** | **Decoder** |
| **Layer Name, Kernel Size, Number of Kernels** | **Layer Name, Kernel Size, Number of Kernels** |
| Conv E_1, $[7 \times 7, \ 64] \times 2$ <br> Max Pooling, $[2 \times 2, \ 1] \times 1$ <br> Conv E_2, $[7 \times 7, 128] \times 2$ <br> Max Pooling, $[2 \times 2, \ 1] \times 1$ <br> Conv E_3, $[7 \times 7, \ 256] \times 2$ <br> Max Pooling, $[2 \times 2, \ 1] \times 1$ <br> Conv E_4, $[7 \times 7, \ 512] \times 2$ <br> Max Pooling, $[2 \times 2, \ 1] \times 1$ <br> Conv E_5, $[7 \times 7, \ 1024] \times 2$ | Upsampling, D_1, $[2 \times 2] \times 1$ <br> Conv D_1, $[7 \times 7, \ 512] \times 2$ <br> Upsampling, D_2, $[2 \times 2] \times 1$ <br> Conv D_2, $[7 \times 7, \ 256] \times 2$ <br> Upsampling, D_3, $[2 \times 2] \times 1$ <br> Conv D_3, $[7 \times 7, \ 128] \times 2$ <br> Upsampling D_4, $[2 \times 2] \times 1$ <br> Conv D_4, $[7 \times 7, \ 64] \times 2$ <br> CNN + Sigmoid, $[7 \times 7, \ 1] \times 1$ |

*Conv = CNN + ReLU

**Supplementary Table S3. Diagnosis of the cases used for Computer-Aided Diagnosis (CAD) accuracy validation.**

| Extent      | Stage     | Grade | The number of cases |
| ----------- | --------- | ----- | ------------------- |
| Localized   | Stage I   | A     | 1                   |
| Localized   | Stage II  | A     | 1                   |
| Localized   | Stage II  | B     | 2                   |
| Generalized | Stage II  | B     | 4                   |
| Localized   | Stage III | B     | 16                  |
| Localized   | Stage III | C     | 10                  |
| Generalized | Stage III | B     | 2                   |
| Generalized | Stage III | C     | 10                  |